\documentclass{SCAE}
\numberwithin{equation}{section}
\begin{document}

\Year{2013} %
\Month{November}
\Vol{56} %
\No{1} %
\BeginPage{1} %
\EndPage{11} %
\AuthorMark{Z. Gan {\it et al.}}
\ReceivedDay{April 23, 2013}
\AcceptedDay{November 12, 2013}
\PublishedOnlineDay{; published online January 22, 2014}
\DOI{10.1007/s11425-000-0000-0} 

\title{Efficient Implementation of the Barnes-Hut Octree Algorithm for Monte Carlo Simulations of Charged Systems}{}


\author[1]{Zecheng Gan}{}
\author[1]{Zhenli Xu}{Corresponding author}

\address[{\rm1}]{Department of Mathematics, Institute of Natural Sciences, and MoE Key Lab
of Scientific and Engineering Computing,\\ Shanghai Jiao Tong University,
Shanghai 200240, China}
\Emails{ganzecheng1988@sjtu.edu.cn,
xuzl@sjtu.edu.cn}\maketitle


 {\begin{center}
\parbox{14.5cm}{\begin{abstract}
 Computer simulation with Monte Carlo is an important tool to investigate the
function and equilibrium properties of many systems with biological and soft
matter materials solvable in solvents. The appropriate treatment of
long-range electrostatic interaction is essential for these charged systems,
but remains a challenging problem for large-scale simulations. We have
developed an efficient Barnes-Hut treecode algorithm for electrostatic
evaluation in Monte Carlo simulations of Coulomb many-body systems. The
algorithm is based on a divide-and-conquer strategy and fast update of the
octree data structure in each trial move through a local adjustment
procedure. We test the accuracy of the tree algorithm, and use it to perform computer simulations of electric double layer near a spherical interface. It
has been shown that the computational cost of the Monte Carlo method with
treecode acceleration scales as $\log N$ in each move. For a typical system
with ten thousand particles, by using the new algorithm, the speed has been
improved by two orders of magnitude from the direct summation.\vspace{-3mm}
\end{abstract}}\end{center}}

 \keywords{Electrostatics, Monte Carlo, Fast algorithms, Octree, Colloidal interfaces.}

 \MSC{41A58, 82D15, 68P05}

\renewcommand{\baselinestretch}{1.2}
\begin{center} \renewcommand{\arraystretch}{1.5}
{\begin{tabular}{lp{0.8\textwidth}} \hline \scriptsize
{\bf Citation:}\!\!\!\!&\scriptsize Z. Gan, Z. Xu, Science  China: Mathematics  title. Sci China Math, 2013, 56, doi: 10.1007/s11425-000-0000-0\vspace{1mm}
\\
\hline
\end{tabular}}\end{center}

\baselineskip 11pt\parindent=10.8pt  \wuhao
\section{Introduction}

Monte Carlo methods are often used in computer simulations of coarse-grained
biological and soft matter materials in aqueous solution, such as in
studying the properties of membranes, DNAs and colloidal suspensions \cite%
{FS:book:02,AT:book:87}. In these simulations, a bottleneck for high
performance computing is to fast and accurately evaluate pairwise
electrostatic interactions in the system. For a group of $N$ charges $%
\{q_1e,q_2e,...,q_Ne\}$ at $\{\mathbf{r}_1,\mathbf{r}_2,...,\mathbf{r}_N \}$,
the total electrostatic energy of the system, $V$, can be expressed as,
\begin{equation}
\beta V=l_B\sum_{i=1}^{N-1}\sum_{j=i+1}^N\frac{q_{i}q_{j}}{|\mathbf{r}_{i}-%
\mathbf{r}_{j}|},
\end{equation}%
where $\beta$ and $l_B$ are two constants representing the inverse thermal
energy and the Bjerrum length of the medium, respectively. A direct summation of this
energy scales as $O(N^2)$ so it becomes prohibitively expensive if large
systems are considered.

A lot of fast algorithms have been developed to speed up this computation.
Due to the long-range nature, simply truncating the system to make a finite
simulation volume is never allowed because of strong artifact effect.
Periodic boundary conditions could reduce this effect and maintain the
particles in the volume but include an infinite series of periodic images.
The Ewald summation \cite{Ewald:AP:21} splits the pairwise interaction into a short-range interaction and a long-range interaction which can be
performed in the physical space and in the Fourier space, respectively, and
a balance of the computation in the two spaces yields an $O(N^{3/2})$
algorithm. This can be further simplified to $O(N\log N)$ when a lattice
summation based on fast Fourier transform is utilized, resulting in
particle-mesh Ewald \cite{DYP:JCP:93} and particle-particle particle-mesh
Ewald \cite{HE:book:88} algorithms, popularly employed in molecular dynamics
packages. Frequently it has been widely interested to simulate nonperiodic systems.
In particular, when a highly charged object is present in the system,
periodic boundary conditions have to use very large simulation volume to
reduce periodic artifacts. Reaction field models with spherical cavity \cite%
{XC:SIREV:11} have been proposed, which include finite number of images to
represent the dielectric effects of the bulk solvent due to the cavity.
Since the particle systems composed of source and image charges are
extremely nonuniform, Barnes-Hut treecode \cite{BH:Nature:86,WSH:CPC:12,GSK:PS:10,MG:JCP:05} and fast
multipole \cite{GR:JCP:87,LC:HPCN:09,KD:FMLRICS:11} algorithms will be beneficial to simulate such
systems.

The treecode algorithms originated from Appel \cite{Appel:SISC:85} and
Barnes and Hut \cite{BH:Nature:86} in the field of astrophysics. Appel used
a binary tree structure in which the clusters can have arbitrary shapes,
while Barnes and Hut used an octree data structure where all clusters are
cubic boxes. These algorithms divide the particles into a hierarchical tree
structure of clusters, and compute the particle-cluster interaction by using
the multipole expansion of the Green's function along the tree. The
Barnes-Hut method and its generalizations have been often used in practical
simulations due to its conceptually simple and easy-to-implement tree
construction. Treecode algorithms have an $O(N\log N)$ computational
complexity for pairwise interaction of $N$ particles and they can be applied
to both periodic and non-periodic boundary conditions \cite%
{DK:JCP:00,DK:JCC:01,LK:JCP:01}, and also Yukawa \cite{LJK:JCP:09} and
generalized Born \cite{XCY:JCP:11} potentials. Additional step by converting
them into local expansions leads to the fast multipole algorithms \cite%
{GR:AN:97,CGR:JCP:99,YBZ:JCP:04} (see Ying \cite{Ying:SCM:12} for a tutorial
review), which could improve the complexity up to $O(N)$.

In Monte Carlo (MC) simulations, the algorithm for electrostatics module
should be specially made. Different from the collective motion of all
particles in dynamic simulations, one-particle displacement is preferred in
each MC step in order to gain a better sampling efficiency \cite{FS:book:02}%
. In this displacement strategy, only particle interactions related to the
chosen particle are required to calculate. Suppose the $i$-th particle is
chosen, then the energy is, $\beta V_i= l_B\sum q_iq_j/r_{ij}, $ a summation
over index $j\neq i$. This expression is calculated twice to compare the
energy difference before and after the displacement of particle $i$, which
is used to decide if the new configuration is accepted or rejected.
Obviously, a simple direct evaluation costs $O(N)$ complexity, and the cost
of using Ewald summation will be $O(N^{1/2})$. The lattice summations are
not advantaged in one-particle-displacement MC simulations. This motivates
the present work to develop an $O(\log N)$ calculation for each MC
displacement with the Barnes-Hut-type algorithm targeting at large scale
simulations of nonuniformly distributed charged systems.

Our algorithm for constructing an efficient electrostatic module of MC
simulations is based on the following intuitive observation. For a given
octree structure, the operations of calculating $V_i$ in each MC step will
be $O(\log N)$ if the multipole expansion is applied, and therefore we
should also make a local modification of the tree structure within $O(\log
N) $ in every particle displacement. This can be guaranteed if the local
adjustment of the octree data is only done for clusters related to the
chosen particle. By this strategy, we test the performance of the treecode
algorithm in simulating a typical charged system in colloidal suspensions,
which demonstrates its attractive feature for a wider application in related
complex biological or soft matter systems.

\section{Primitive model of charged systems and Monte Carlo simulations}

\begin{figure}
\includegraphics[width=0.6\textwidth]{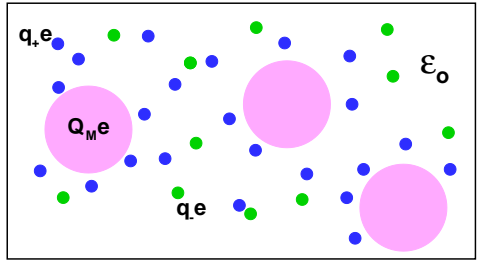}
\caption{Schematic illustration of the primitive model of electrolytes, including spherical macroions (pink) with valence $Q_M$, counterions (blue) and coions (green) with valence $q_+$ and $q_-$ respectively. These particles at different scales are mobile in a solvent which affects the ions by its dielectric permittivity $\varepsilon_{\mathrm{o}}$.}
\label{pri}
\end{figure}

Let us start with a typical charged system for colloidal suspensions composed of macroions and small ions, schematically shown in Fig. \ref{pri}. One common concern in different disciplines is the ionic structure near a charged surface, which is the so-called electric double layer (EDL) \cite{FPPR:RMP:10}, since in the EDL ions are strongly correlated and ion distribution can not be correctly predicted by mean-field theory. The primitive model is often used to describe EDLs  \cite{Linse:APS:05,WKDG:NS:11}, because it takes into account ion-ion interactions explicitly. Within the primitive model framework, ionic species are represented by charged hard spheres of different sizes with charges at spherical centers. The solvent is treated implicitly and only enters the model with its relative permittivity $\varepsilon_\mathrm{o}$. The Hamiltonian of an EDL system can be written as a pairwise summation over all macro- and small ions, $V = \sum_{i<j}V_{ij},  $
and $V_{ij}$ is the superposition of the electrostatic interaction and the hard-sphere repulsion between two particles,
\begin{equation}
\beta V_{ij}=\left\{\begin{array}{ll}
\frac{l_Bq_{i}q_{j}}{r_{ij}},~~\hbox{if}~~r_{ij}>(\tau_i+\tau_j)/2,\\
~~\infty,~~~~\hbox{if}~~r_{ij}\leq (\tau_i+\tau_j)/2,
\end{array}\right.
\end{equation}
where $\beta=1/k_BT$ is the inverse thermal energy, $l_B=e^2/4\pi\varepsilon_0\varepsilon_\mathrm{o} k_BT$ is the Bjerrum length, $\varepsilon_0$ is the vacuum permittivity and $\varepsilon_\mathrm{o}$ is the relative permittivity, $q_i$ is the valence, $\tau_i$ is the diameter of ion,  and $r_{ij}$ is the distance between two ions.

In equilibrium, the system follows the Boltzmann distribution $P(X)=e^{-\beta V(X)}/Z$, where $X$ is a sample in a high dimensional space ($3N$ dimensions where $N$ is the number of particles) and presents a possible configuration of the system, and $Z=\int_\Omega e^{-\beta V(X)}dX$ is a normalization constant.  The partition function $Z$ is an integral over the whole sample space $\Omega$.  The Monte Carlo (MC) algorithm \cite{FS:book:02} is a computational
approach to obtain equilibrium distribution of such a system. It
samples $M$ configurations in $\Omega$ by a Markov
chain, ${X_1,X_2,...,X_M}$, such that for any configuration $X$  the following limit is satisfied,
\begin{equation}
\lim_{M\rightarrow \infty}\frac{M_X}{M}=P(X),
\end{equation}
where $M_X$ is the number of configurations $X$ out of the sampled $M$ configurations.

In general, the Markov chain is constructed through the Metropolis principle \cite{Metropolis:53}. The implementation of the Metropolis MC algorithm can be divided into the following steps:
\begin{enumerate}
\item Choose an initial configuration $X_1$ from the sample space. This can be done by randomly placing particles in simulation volume. Set index $m=1$.
\item  Choose a trial configuration $X_t$, compute the ratio $\Theta= P(X_t)/P(X_m)=e^{-\beta[V(X_t)-V(X_m)]}$, and generate a random number $\alpha$ from $[0,1]$. Set the next configuration of the Markov chain, $X_{m+1}$=$X_t$ if $\alpha \leq \Theta$, otherwise copy the old configuration as the new configuration $X_{m+1}=X_m$.
\item Let index $m=m+1$ and go to step 2 for next configuration.
\end{enumerate}
With the Markov chain, any equilibrium property of the system, $\langle B\rangle$, which is an ensemble average, can be calculated through,
\begin{equation}
\langle B\rangle\thickapprox \frac{1}{M}\sum_{X}M_{X}B(X).
\end{equation}
Here $B(X)$ is the value of $B$ for a state $X$.

The benefit of the Metropolis algorithm is to avoid computing $Z$ directly, and instead only energy difference of two configurations $ V(X_t)-V(X_m) $ is needed to compute, which is the most expensive part of the algorithm. Another benefit, we can see from Step 2, is that the algorithm allows a flexible choice of a candidate configuration which should satisfy the detailed balance condition, $P(X)p(X\rightarrow Y)=P(Y)p(Y\rightarrow X)$, where $p$ is the transition probability. A suitable choice of the candidate configuration will greatly increase the sampling efficiency. In consideration of computing the energy difference in a low price, Metropolis MC algorithm usually displaces one randomly chosen particle to generate a new configuration in each step \cite{FS:book:02}, which has similar sampling efficiency as displacing multiple particles for a given acceptance ratio, but greatly improves the computation efficiency of the electrostatic energy.

\section{Monte Carlo implementation of the tree algorithm}

\subsection{Multipole expansion and treecode algorithm}

\begin{figure}[tbph]
\includegraphics[width=0.6\textwidth]{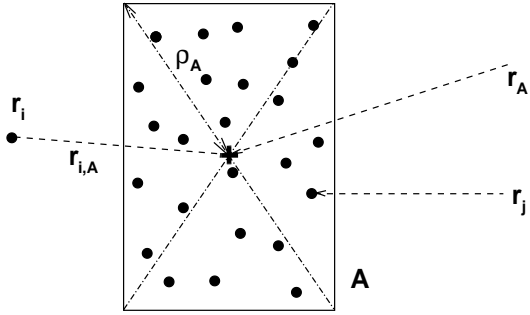}
\caption{Illustration of interaction between a particle $i$ and particles in
a distant cluster $A =\{\mathbf{r}_j, j=1,2,..., N_A\}$. $\mathbf{r}_{i,A}$ is
the displacement between the particle to the cluster center, $\mathbf{r}_A$ is
the center of the cluster, and $\protect\rho_A$ is the radius of the
cluster. }
\label{pci}
\end{figure}

In this section, we overview the mathematical formulation  of the treecode
algorithm which will be used for energy calculation, i.e., the multipole approximation for the particle-cluster
interaction with truncating multivariate Taylor and the recurrence relations
for the expansion coefficients. In Cartesian coordinates, $\mathbf{r}=(x_1,
x_2, x_3)$, the multivariate Taylor expansion of the Green's function $f(\mathbf{r})=1/|\mathbf{r}%
|$ at $\mathbf{r}^{\prime }$ can be written as,
\begin{equation}
f(\mathbf{r})=\sum_{||\mathbf{n}||=0 }^{\infty }\frac{1}{\mathbf{n}!}D_{\mathbf{r}}^{\mathbf{n}}f(\mathbf{r}^{\prime })(\mathbf{r-r}^{\prime })^\mathbf{n},  \label{taylor}
\end{equation}%
where the multi-index notation $||\mathbf{n}||= n_1+n_2+n_3$, $\mathbf{n}!= n_1!n_2!n_3!$,
$D_{\mathbf{x}}^{\mathbf{n}}=\partial ^{||\mathbf{n||}}/(\partial x_1^{n_1}\partial x_2^{n_2}\partial x_3^{n_3})$,
and $(\mathbf{x-x}')^{\mathbf{n}}=(x_1-x_1')^{n_1}(x_2-x_2')^{n_2}(x_3-x_3')^{n_3}$. We
will use this expression for the multipole approximation of a particle $i$
and particles in a distant cluster $A$ which has a center $\mathbf{r}_A$
(shown in Fig. \ref{pci}). The electrostatic energy between particle $i$ and
particles in the cluster $A$ is,
\begin{equation}
\beta V_{i,A}=l_B\sum_{j\in A}\frac{q_{i}q_{j}}{|\mathbf{r}_{j}-\mathbf{r}%
_{i}|}.  \label{eng}
\end{equation}
Substituting Eq. \eqref{taylor} into Eq. \eqref{eng} by setting $\mathbf{r}=%
\mathbf{r}_j-\mathbf{r}_i$ and $\mathbf{r}^{\prime }=\mathbf{r}_{A}-\mathbf{r%
}_i$, and truncating the expansion at $||\mathbf{n}||=p$, we obtain the $p$%
th order multipole approximation to $V_{i,A}$ as,
\begin{eqnarray}
\beta V_{i,A} &=&l_B\sum_{j\in A}q_iq_jf(\mathbf{r}_j-\mathbf{r}_i)  \notag \\
&\approx &l_B q_i\sum_{||\mathbf{n}||=0}^{p}\frac{1}{\mathbf{n!}}D_{\mathbf{r}}^{%
\mathbf{n}}f(\mathbf{r}_A-\mathbf{r}_i)\sum_{j\in A}q_j(\mathbf{r}_j-\mathbf{%
r}_A)^{\mathbf{n}}  \notag \\
&=&l_Bq_i\sum_{||\mathbf{n}||=0}^{p}T_{\mathbf{n}}(\mathbf{r}_A-\mathbf{r}%
_i)M_A^{\mathbf{n}},  \label{exp}
\end{eqnarray}%
where $T_{\mathbf{n}}(\mathbf{r}_A-\mathbf{r}_i)=\frac{1}{\mathbf{n!}}D_{%
\mathbf{r}}^{\mathbf{n}}f(\mathbf{r}_A-\mathbf{r}_i)$ is the $\mathbf{n}$th
order coefficient of the multidimensional Taylor expansion and $M_A^{\mathbf{n}}=\sum_{j\in
A}q_j(\mathbf{r}_j-\mathbf{r}_A)^{\mathbf{n}}$ is the corresponding
multipole moment of cluster $A$. It should be noted that the moment $M_A^%
\mathbf{n}$ is independent of the distant particle $i$, and thus it can be
pre-calculated once for the use of alternating the distant particle $i$.
Another important observation is that the coefficient $T_{\mathbf{n}}(%
\mathbf{r}_{A}-\mathbf{r}_i)$ only depends on the displacement between the
particle $i$ and the cluster center and is independent of particles in the
cluster, and the calculation of this quantity is supposed to be efficient in
simulations.

The Taylor coefficients $T_\mathbf{n}$ can be calculated by using recurrence
relations \cite{DK:JCC:01},
\begin{equation}
||\mathbf{n }||  r^2T_{\mathbf{n}}(\mathbf{r})+(2||\mathbf{%
n||-}1)\sum_{i=1}^{3}x_{i}T_{\mathbf{n-e}_{i}}(\mathbf{r})+(||\mathbf{n}||-1
)\sum_{i=1}^{3}T_{\mathbf{n-2e}_{i}}(\mathbf{r})=0,
\end{equation}%
where $r=|\mathbf{r}|.$
Since the Taylor coefficient $T_{\mathbf{n}}(\mathbf{r})=0$ when a
certain index $n_i$ is negative, these recurrence relations allow us to efficiently calculate expansion coefficients in the multipole approximation. The error of the $p$th order multipole approximation between particle $i$ and a particle at $\mathbf{r}_j$
inside cluster $A$ can be estimated by bounding
the sum of the $n>p$ terms, which is given by \cite{DK:JCC:01},
\begin{eqnarray}
 \sum_{n=p+1}^{\infty }\sum_{||\mathbf{n||=}n}^{{}}\left\vert T_{%
\mathbf{n}}(\mathbf{r}_A-\mathbf{r}_i)(\mathbf{r}_j-\mathbf{r}_A)^{\mathbf{n}%
}\right\vert &\leq &\frac{1}{R} \left( \frac{\gamma
^{p+1}}{1-\gamma }\right) .  \label{err}
\end{eqnarray}%
where $\rho _A$ is the radius of the cluster equal to half length of the box diagonal,  $R=r_{i,A}$ and $\gamma= \rho_A/R$ should be less than 1. We can
see increasing either $R$ or the expansion order $p$ could improve the accuracy of the multipole expansion.  In practice, the error could be even much smaller due to the error cancellation of different particles in $A$, in particular, when the particle distribution is uniform.

\subsection{Calculation of electrostatic energy}

The calculation of electrostatic energy in the Hamiltonian of charged systems is
based on a hierarchical octree of the data structure. Each node in the tree is
a cluster of particles. One of them is the root cluster, which is a cubic
box enclosing the whole simulation domain (usually the smallest box
including all the particles is taken), and we say it is the zeroth level of
the tree structure. In our algorithm, the root cluster is bisected in each
coordinate direction to yield eight children clusters which form the first
level of the tree. This process is applied recursively on the children
clusters until a user-specified level $L_\mathrm{max}$ is reached. The
clusters at level $L_\mathrm{max}$ are also called leaves, and $L_\mathrm{max%
}$ logarithmically depends on the number of particles in the system.

The generation of the tree structure is order $N\log N$ algorithm. Once the
tree has been constructed, it will remain a fixed structure throughout the
MC simulation and only a local modification is allowed. In the data
structure, each node, i.e., a certain cluster $A$, includes the following
information of $A$:

\begin{enumerate}
\item The number of particles in cluster $A$, $N_A$. For example, the root
cluster has $N$ particles.

\item The radius of the cluster $\rho_A$, which is equal to a half of the
box diagonal length.

\item The minimal, middle and maximal values of the $x,y,z$ coordinates of
the cluster box.

\item A value, $l_A$, to indicate at which level the cluster is located.

\item An index array which stores all particles in the cluster.

\item All the moments for the truncated multipole expansion.

\item The pointers to its eight children if $A$ is not a leaf cluster.
\end{enumerate}

The potential energy of a particle $i$ interacting with other particles is
the summation of particle-cluster interactions by the multipole
approximation or particle-particle interactions by the direct summation. It
is calculated by traversing the tree recursively, starting from the root
cluster. If the particle-cluster interaction being employed is determined
according to the multipole acceptance criterion (MAC), $\rho_A/r_{i, A}\leq
\theta$, and if $N_A\geq N_0$ is satisfied, where $\theta\in[0,1]$ and the
minimum expansion number $N_0$ are two user-specified parameters. If both
two conditions are satisfied, the multipole approximation is used. If the
MAC is not satisfied, but $N_A>N_0$ and $A$ is not a leaf cluster, then the
eight children of $A$ will be examined, because the radius of the children
clusters will be smaller, the MAC is more likely to be satisfied. For all
the other cases, the direct summation method is used.

\subsection{Local modification of the tree structure}

Due to the possible position change of the chosen particle, a minor
modification to the data information of the tree structure should be done in
each MC displacement. In order to guarantee that the local adjustment
procedure can be accomplished in $O(\log N)$ operations, we need to use the
tracking information for all the particles, which is stored initially. At a
certain level we know the cluster including this particle, the tracking
information of this particle essentially remembers the location at the index
array of this particle. This information is useful in reducing the
operations.

\begin{figure*}[tbph]
\begin{center}
\includegraphics[width=0.8\textwidth]{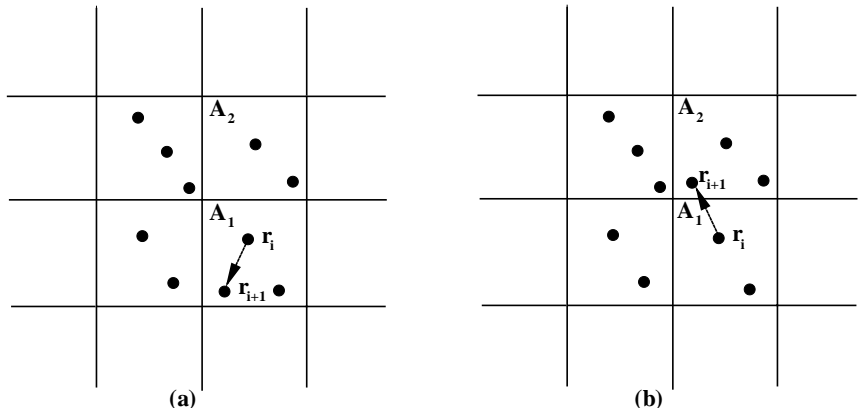}
\end{center}
\caption{Two possible cases which could happen for an MC trial move. At the
same level, (a) the particle $i$ stays inside a cluster $A_1$; (b) it moves
from $A_1$ to another cluster $A_2.$ }
\label{move}
\end{figure*}

Suppose a trial move takes place during an MC step of trying to move the $i$%
th particle. At each level of the octree, there could be two possible cases
(Fig. \ref{move}): (a) Particle $i$ stays inside the original cluster $A_1$, or (b)
it moves from $A_1$ to another cluster $A_2$. Suppose case (a) happens, then
we only need to update the moments stored in the node of $A_1$. If it is
case (b), then we need an ``add" operation to $A_2$ and a ``delete"
operation to $A_1$. In the add operation, we need to update $N_{A_2}$, the
moments by adding the contribution from $i$, and the index array adding the
index $i$ at the end of this index array, and finally the tracking
information of $i$. In the delete operation, we update $N_{A_1}$ and the
moments by extracting the contribution from $i$, and in the index array, we
replace $i$th index by the one at the end of the array and update the
tracking information of $k$. Both the add and delete operations are
performed recursively.

The new MC algorithm with treecode acceleration is summarized as follows.
Initially, the locations and charges of all particles are generated, and the
parameters such as the MAC parameter $\theta$, the order of the multipole
expansion $p$, and the maximal expansion number $N_{0}$ are prescribed. Then
the octree structure and the tracking information for each particle are
built. At every MC step, we calculate the electrostatic energy with $O(N)$
operations. The cell-list algorithm is used for the short range interaction
between particles, which has $O(1)$ operations. The local adjustment
procedure is then performed to update the data information in the tree. If
the trial move is rejected, the tree structure remains unchanged.

\section{Numerical Results}

In order to test the performance of the algorithm, we have implemented the Metropolis MC simulation to calculate the equilibrium structure of the spherical EDLs. The EDL is composed of positively and negatively charged ions surrounding a macroion, all ions are of spherical shapes. The spherical cell model \cite{Linse:APS:05} is used to define the simulation volume, which confines all ions in a spherical cell, i.e., if any ion moves across the cell boundary in an MC step, the trial move will be rejected by the hard boundary. The macroion is fixed and concentric with the spherical cell, while all the small ions can move freely inside the cell. Due to its importance in colloidal science, biophysics, and electrochemical energy and so on, computer simulations of the EDL structure have attracted a wide interest \cite{Linse:JCP:08,MMM:TJPC:93,DJHL:TJPCB:01,LAT:PRL:98,KGN:PCCP:11}.

In the calculations, we take the Bjerrum length $l_B=7.14 \AA$, which represents the water solvent at room temperature. Besides the macroion, only counterions are present in each system, which have a uniform diameter $\tau=4 \AA$ and positive charge $q=2$ elementary charges. The radius of the macroion
takes $a=30 \AA$, which is at the same scale as a charged surfactant micelle \cite{Linse:APS:05}. The radius of the simulation cell is $R_\mathrm{cell}=80 \AA$. In testing the timing, the number of counterions, $N$, is varied from $10$ to $10k$, the valence of the macroion takes $Q_M = -qN$ to guarantee the electroneutrality of each system.

In electrostatic algorithm based on the Barnes-Hut tree structure,  the MAC parameter takes two values of $\theta =0.5$ and 0.3 for comparison, and the order of the multipole expansion $p$ varies between 0 and 10. The minimal expansion number $N_0$ is optimized as a function of $p$ by pre-computing execution times, which is determined to be the smallest number of particles in a cluster such that the multipole approximation is more efficient than the direct summation. In each simulation, $10^6N$ MC trial moves are attempted for the convergence of the empirical distribution.  The computations were performed on a Linux machine with 2.67 GHz CPU and 48 G memory.

\begin{figure*}[tbph]
\begin{center}
\includegraphics[width=0.45\textwidth]{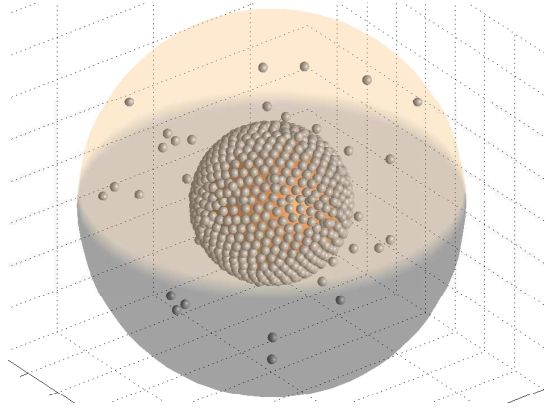}
\end{center}
\caption{A snapshot configuration of charge distribution around a macroion with 800 counterions.} \label{sedl}
\end{figure*}

\begin{figure*}[tbph]
\includegraphics[width=0.45\textwidth]{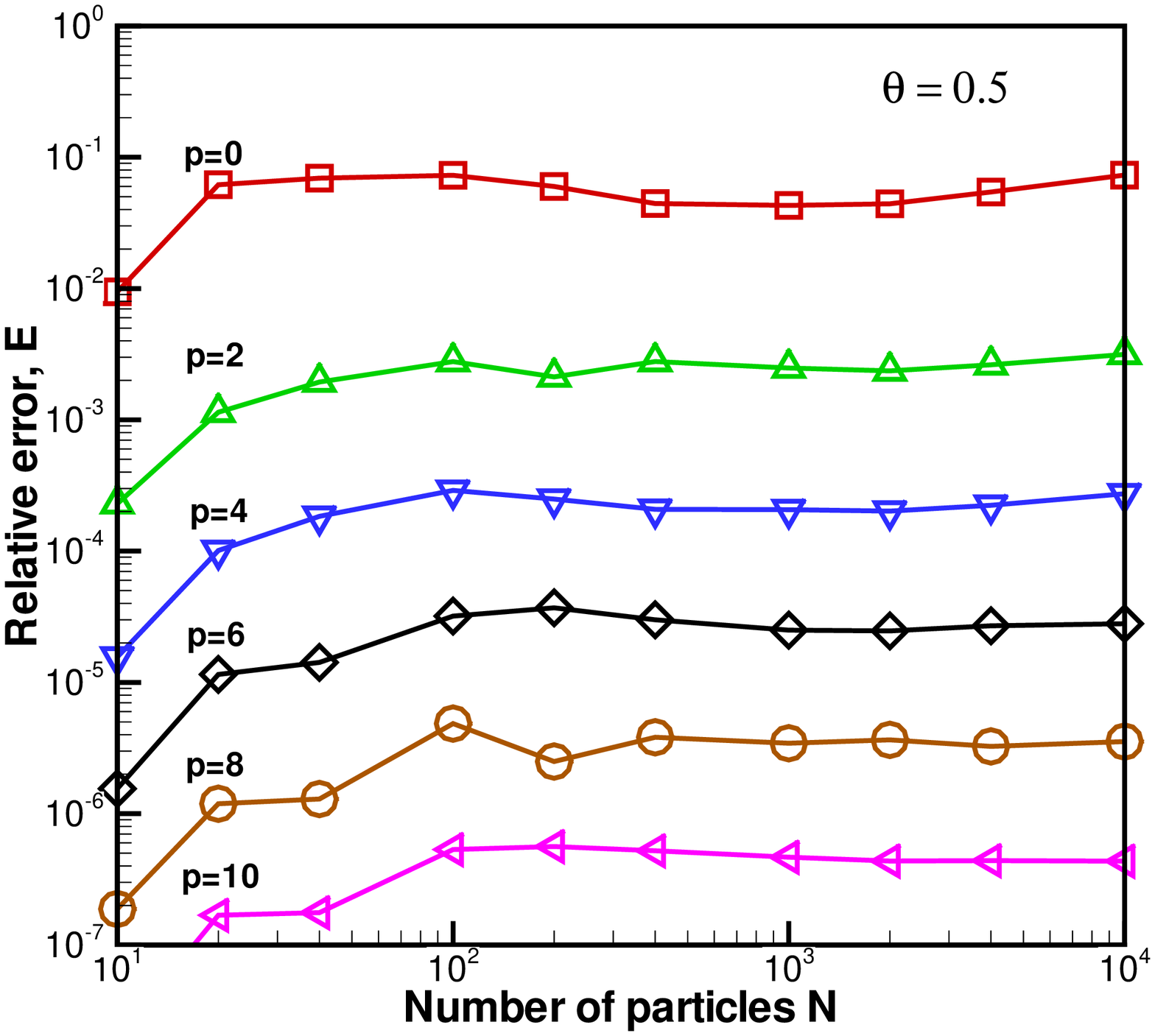}
\includegraphics[width=0.45\textwidth]{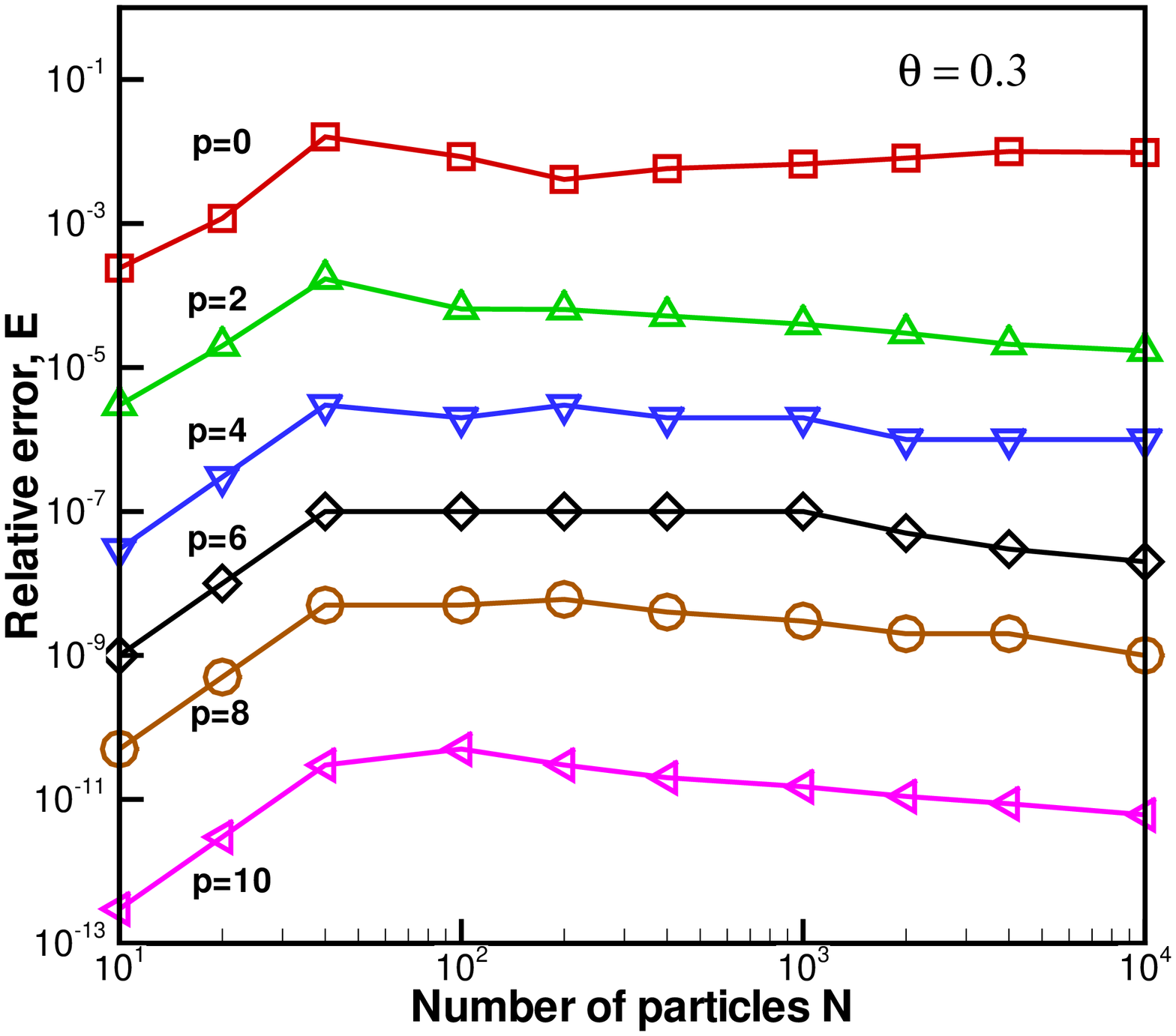}
\caption{Relative errors in electrostatic potential energy, $E$, computed by the treecode algorithm with multipole approximation order $p$ from 0 to 10. (a) $\theta =0.5$; and (b) $\theta =0.3$. } \label{error}
\end{figure*}

In order to verify the accuracy of the algorithm, we randomly choose a snapshot configuration of the spherical EDL with $N=$ $10$ to $10\mathrm{k}$, and calculate the error of different orders based on this configuration. A typical configuration with $N=$ $800$ is illustrated in Fig. \ref{sedl}, where we can see the charge distribution is not uniform and most mobile ions are aggregated near the surface. Although the strong surface charge is not physical, this acts as a good test for the performance of the multipole expansion since there is no error cancellation for charges in a cluster having the same sign. We measure the relative error $E$ of the treecode algorithm, which
is defined by,
\begin{equation}
E=\left(\frac{\sum_{i=1}^{N}|V_{i}-\overline{V_{i}}|^{2}}{%
\sum_{i=1}^{N}|V_{i}|^{2}}\right)^{1/2},
\end{equation}
where $V_i$ is the exact potential on particle $i$ obtained by the direct summation method and $\overline{V_i}$ is the potential obtained using
treecode algorithm. The results are shown in Fig. \ref{error}. We can see that when $N\geq$ $100$, the error is almost independent of the system size $N.$  The relative error decreases with the increase of order $p.$ Although most counterions assemble near the macroion surface and the system is not uniform, when $p=2$ the error is already less than $1\%$ for $\theta=0.5$ and $0.1\%$ for $\theta=0.3$.

\begin{figure*}[tbph]
\includegraphics[width=0.45\textwidth]{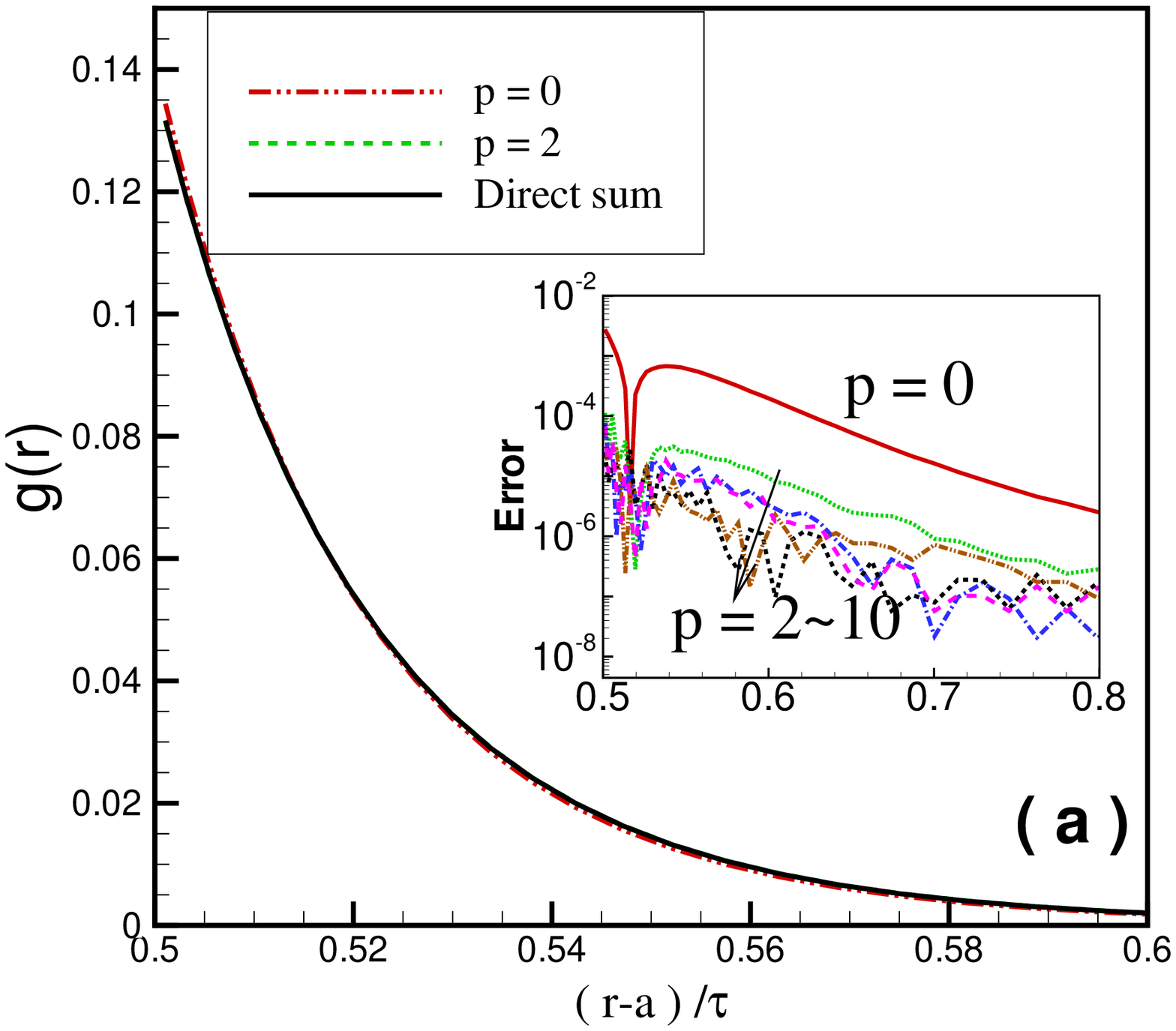}\includegraphics[width=0.48\textwidth]{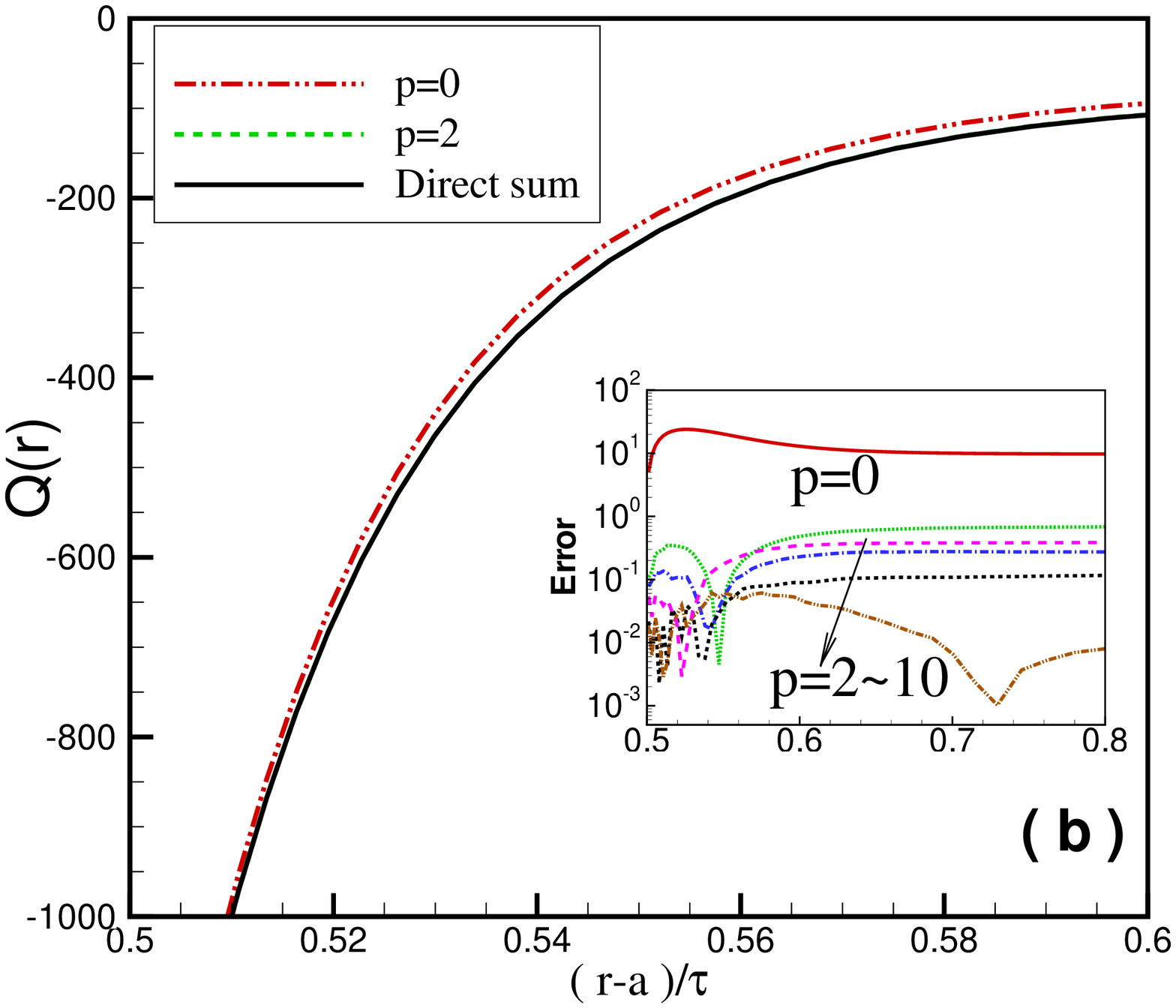}
\caption{Radial distribution functions (a) and integrated charge distribution functions (b) of a typical system with $N=800$, computed by using direct summation, and multipole approximations with $\theta=0.5$ and order $p$ from $0$ to $10$. The curves of $p=2\sim10$ and the direct summation are overlapping, and the errors deviated from the direct summation can be found from the inset plots.} \label{rdf}
\end{figure*}

Using Monte Carlo simulations, we also test the accuracy by investigating two physical properties of double layer structure: the radial distribution
function (RDF), $g(r)$, of the counterions and the integrated charge distribution function (ICDF), $Q(r)$, of the systems. In simulations, it is calculated using the average particle in a bin $N(r,r+\Delta r)$ divided the volume of the bin and normalized the sum of all bins. The RDF describes the normalized average density along the radial direction. And the ICDF $Q(r)$ is a function of $r$, which represents the average charge with the sphere of radius $r$. Clearly, we have $Q(a)=Q_M$ and $Q(R_\mathrm{cell})=0$. To observe the accuracy of these two functions, we plot the simulation results of $N=800$ and $\theta=0.5$ in Fig. \ref{rdf}. We see both functions converge to the results of direct summation,  starting from $p=2$. Shown in the inset plots, the errors are substantially in the same order as $p\geq 2$ due to the fluctuation of the MC sampling, demonstrating that the multipole expansion with $p=2$ provides enough accuracy for describing the physical properties of the studied system. It can be observed that both $g(r)$ and $Q(r)$ exponentially decay near the charged interface, in agreement with the prediction from the strong-coupling theory \cite{BKNNSS:PP:05}. In contrast, the decay rate predicted by the classical Poisson-Boltzmann theory is algebraic \cite{BKNNSS:PP:05}, and thus these simulation results demonstrate the the theory fails for strong-coupling charged systems \cite{GNS:RMP:02,LLPS:PRE:02}.

\begin{figure*}[tbph]
\begin{center}
\includegraphics[width=0.6\textwidth]{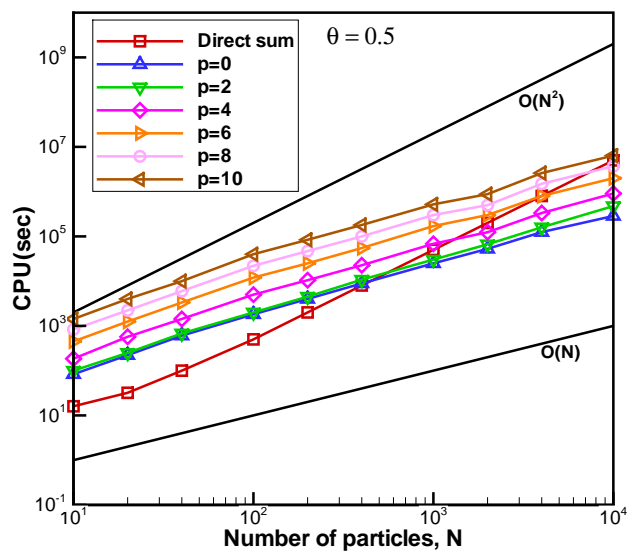}
\end{center}
\caption{ CPU times in MC simulations of the spherical EDLs with the direct summation and treecode acceleration for $p=0\sim10$ as a function of particle number $N$.} \label{time}
\end{figure*}

Finally, the CPU times of the MC algorithm are compared for different strategies of the electrostatic calculations, with different
system size $N$ and expansion order $p$. The results have been shown in Fig. \ref{time}. Since the number of MC steps is of $O(N)$, it is illustrated that the CPU times for those curves with
treecode acceleration scale as $O(N\log N)$ while  that of direct summation scales as $O(N^2)$. The case of $p=2$, which is accurate enough to provide correct RDF and ICDF curves, has the break-even point about $N=600.$ Using the treecode acceleration, we find much larger systems can be explored, and the simulation for a system of $N=10\mathrm{k}$ can be done in a few days using the new MC algorithm, which is around $100$ times faster than that of the direct summation.

\section{conclusion}
In this paper, we have developed an efficient generalization of the Barnes-Hut octree algorithm in Monte Carlo simulations of charged systems. In order to adapt the one-particle displacement of Metropolis Monte Carlo simulations, the algorithm introduces a local modification strategy of the tree structure, which leads to a low computational cost, namely, $O(\log N)$ complexity in each Monte Carlo step. This allows us to simulate equilibrium properties of a system with more than ten thousand charged particles in personal computers.

Future goals of this project includes the parallelization of the treecode algorithm \cite{WSH:CPC:12,GK:JCP:13}, the incorporation into the multiscale reaction-field modeling of electrolytes and with image charges for colloidal suspensions \cite{GX:PRE:11,XLX:SIAP:13}, and the exploration of properties of larger scaled systems.

\Acknowledgements{The authors acknowledge the financial support from the Natural Science
Foundation of China (Grants No.: 11101276 and 91130012). Z. X. acknowledges the support from the Alexander
von Humboldt foundation for a research stay at the ICP, University of Stuttgart.}



\begin{thebibliography}{99}
\bahao\baselineskip 11.5pt

\bibitem{AT:book:87}
{\sc Allen, M.~P., and Tildesley, D.~J.}
\newblock {\em Computer Simulations of Liquids}.
\newblock Oxford University Press, Oxford, 1987.

\bibitem{Appel:SISC:85}
{\sc Appel, A.}
\newblock An efficient program for many-body simulations.
\newblock {\em SIAM J. Sci. Stat. Comput. 6\/} (1985), 85--103.

\bibitem{BH:Nature:86}
{\sc Barnes, J., and Hut, P.}
\newblock A hierarchical {O(NlogN)} force-calculation algorithm.
\newblock {\em Nature 324\/} (1986), 446--449.

\bibitem{BKNNSS:PP:05}
{\sc Boroudjerdi, H., Kim, Y.-W., Naji, A., Netz, R.~R., Schlagberger, X., and
  Serr, A.}
\newblock Statics and dynamics of strongly charged soft matter.
\newblock {\em Phys. Rep. 416\/} (2005), 129--199.

\bibitem{CGR:JCP:99}
{\sc Cheng, H., Greengard, L., and Rokhlin, V.}
\newblock A fast adaptive multipole algorithm in three dimensions.
\newblock {\em J. Comput. Phys. 155\/} (1999), 468--498.

\bibitem{DYP:JCP:93}
{\sc Darden, T.~A., York, D.~M., and Pedersen, L.~G.}
\newblock {Particle mesh Ewald: an Nlog(N) method for Ewald sums in large
  systems}.
\newblock {\em J. Chem. Phys. 98\/} (1993), 10089--10092.

\bibitem{DJHL:TJPCB:01}
{\sc Deserno, M., Jim¨¦nez-¨¢ngeles, F., Holm, C., and Lozada-Cassou, M.}
\newblock Overcharging of dna in the presence of salt: Theory and simulation.
\newblock {\em J. Phys. Chem. B 105\/} (2001), 10983--10991.

\bibitem{DK:JCP:00}
{\sc Duan, Z.~H., and Krasny, R.}
\newblock An {Ewald} summation based multipole method.
\newblock {\em J. Chem. Phys. 113\/} (2000), 3492--3495.

\bibitem{DK:JCC:01}
{\sc Duan, Z.~H., and Krasny, R.}
\newblock An adaptive treecode for computing nonbonded potential energy in
  classical molecular systems.
\newblock {\em J. Comput. Chem. 22\/} (2001), 184--195.

\bibitem{Ewald:AP:21}
{\sc Ewald, P.~P.}
\newblock Die berechnung optischer und elektrostatischer gitterpotentiale.
\newblock {\em Ann. Phys. 369\/} (1921), 253--287.

\bibitem{FPPR:RMP:10}
{\sc French, R.~H., Parsegian, V.~A., Podgornik, R., Rajter, R.~F., Jagota, A.,
  Luo, J., Asthagiri, D., Chaudhury, M.~K., Chiang, Y.-M., Granick, S.,
  Kalinin, S., Kardar, M., Kjellander, R., Langreth, D.~C., Lewis, J., Lustig,
  S., Wesolowski, D., Wettlaufer, J.~S., Ching, W.-Y., Finnis, M., Houlihan,
  F., von Lilienfeld, O.~A., van Oss, C.~J., and Zemb, T.}
\newblock Long range interactions in nanoscale science.
\newblock {\em Rev. Mod. Phys. 82}, 2 (2010), 1887--1944.

\bibitem{FS:book:02}
{\sc Frenkel, D., and Smit, B.}
\newblock {\em Understanding molecular simulation: From algorithms to
  applications}.
\newblock Academic Press, New York, 2002.

\bibitem{GX:PRE:11}
{\sc Gan, Z., and Xu, Z.}
\newblock Multiple-image treatment of induced charges in {Monte Carlo}
  simulations of electrolytes near a spherical dielectric interface.
\newblock {\em Phys. Rev. E 84\/} (2011), 016705.

\bibitem{GK:JCP:13}
{\sc Geng, W., and Krasny, R.}
\newblock A treecode-accelerated boundary integral {Poisson-Boltzmann} solver
  for electrostatics of solvated biomolecules.
\newblock {\em J. Comput. Phys. 247\/} (2013), 62 -- 78.

\bibitem{GSK:PS:10}
{\sc Gibbon, P., Speck, R., Karmakar, A., Arnold, L., Frings, W., Berberich,
  B., Reiter, D., and Masek, M.}
\newblock Progress in mesh-free plasma simulation with parallel tree codes.
\newblock {\em Plasma Science, IEEE Trans. on 38}, 9 (2010), 2367--2376.

\bibitem{GR:JCP:87}
{\sc Greengard, L., and Rokhlin, V.}
\newblock A fast algorithm for particle simulations.
\newblock {\em J. Comput. Phys. 73\/} (1987), 325--348.

\bibitem{GR:AN:97}
{\sc Greengard, L., and Rokhlin, V.}
\newblock A new version of the {Fast Multipole Method} for the {Laplace}
  equation in three dimensions.
\newblock {\em Acta Numerica 6\/} (1997), 229--269.

\bibitem{GNS:RMP:02}
{\sc Grosberg, A.~Y., Nguyen, T.~T., and Shklovskii, B.~I.}
\newblock Colloquium: The physics of charge inversion in chemical and
  biological systems.
\newblock {\em Rev. Mod. Phys. 74}, 2 (Apr 2002), 329--345.

\bibitem{HE:book:88}
{\sc Hockney, R.~W., and Eastwood, J.~W.}
\newblock {\em Computer simulation using particles}.
\newblock Taylor \& Francis, 1988.

\bibitem{KD:FMLRICS:11}
{\sc Kabadshow, I., and Dachsel, H.}
\newblock The error-controlled fast multipole method for open and periodic
  boundary conditions.
\newblock {\em Fast Methods for Long-Range Interactions in Complex Systems\/}
  (2011), 85.

\bibitem{KGN:PCCP:11}
{\sc Kondrat, S., Georgi, N., Fedorov, M.~V., and Kornyshev, A.~A.}
\newblock A superionic state in nano-porous double-layer capacitors: insights
  from monte carlo simulations.
\newblock {\em Phys. Chem. Chem. Phys. 13\/} (2011), 11359--11366.

\bibitem{LC:HPCN:09}
{\sc Lashuk, I., Chandramowlishwaran, A., Langston, H., Nguyen, T.-A., Sampath,
  R., Shringarpure, A., Vuduc, R., Ying, L., Zorin, D., and Biros, G.}
\newblock A massively parallel adaptive fast-multipole method on heterogeneous
  architectures.
\newblock In {\em High Performance Computing Networking, Storage and Analysis,
  Proceedings of the Conference on\/} (2009), IEEE, pp.~1--12.

\bibitem{LLPS:PRE:02}
{\sc Lau, A. W.~C., Lukatsky, D.~B., Pincus, P., and Safran, S.~A.}
\newblock Charge fluctuations and counterion condensation.
\newblock {\em Phys. Rev. E 65\/} (2002), 051502.

\bibitem{LJK:JCP:09}
{\sc Li, P., Johnston, H., and Krasny, R.}
\newblock A {Cartesian treecode} for screened {Coulomb} interactions.
\newblock {\em J. Comput. Phys. 228\/} (2009), 3858--3868.

\bibitem{LK:JCP:01}
{\sc Lindsay, K., and Krasny, R.}
\newblock A particle method and adaptive treecode for vortex sheet motion in
  {3-D} flow.
\newblock {\em J. Comput. Phys. 172\/} (2001), 879--907.

\bibitem{Linse:APS:05}
{\sc Linse, P.}
\newblock Simulation of charged colloids in solution.
\newblock {\em Adv. Polym. Sci. 185\/} (2005), 111--162.

\bibitem{Linse:JCP:08}
{\sc Linse, P.}
\newblock Electrostatics in the presence of spherical dielectric
  discontinuities.
\newblock {\em J. Chem. Phys. 128\/} (2008), 214505.

\bibitem{LAT:PRL:98}
{\sc Lyubartsev, A.~P., Tang, J.~X., Janmey, P.~A., and Nordenski{\"o}ld, L.}
\newblock Electrostatically induced polyelectrolyte association of rodlike
  virus particles.
\newblock {\em Phys. Rev. Lett. 81\/} (1998), 5465--5468.

\bibitem{MMM:TJPC:93}
{\sc Manzanares, J., Murphy, W., Mafe, S., and Reiss, H.}
\newblock Numerical simulation of the nonequilibrium diffuse double layer in
  ion-exchange membranes.
\newblock {\em J. Phys. Chem. 97\/} (1993), 8524--8530.

\bibitem{MG:JCP:05}
{\sc Marzouk, Y.~M., and Ghoniem, A.~F.}
\newblock K-means clustering for optimal partitioning and dynamic load
  balancing of parallel hierarchical n-body simulations.
\newblock {\em J. Comp. Phys 207}, 2 (2005), 493--528.

\bibitem{Metropolis:53}
{\sc Metropolis, N., Rosenbluth, A.~W., Rosenbluth, M.~N., Teller, A.~H., and
  Teller, E.}
\newblock Equation of state calculations by fast computing machines.
\newblock {\em J. Chem. Phys. 21\/} (1953), 1087.

\bibitem{WKDG:NS:11}
{\sc Walker, D.~A., Kowalczyk, B., {de la Cruz}, M.~O., and Grzybowski, B.~A.}
\newblock Electrostatics at the nanoscale.
\newblock {\em Nanoscale 3\/} (2011), 1316--1344.

\bibitem{WSH:CPC:12}
{\sc Winkel, M., Speck, R., H{\"u}bner, H., Arnold, L., Krause, R., and Gibbon,
  P.}
\newblock A massively parallel, multi-disciplinary {Barnes-Hut} tree code for
  extreme-scale {N-body} simulations.
\newblock {\em Comput. Phys. Comm. 183\/} (2012), 880--889.

\bibitem{XC:SIREV:11}
{\sc Xu, Z., and Cai, W.}
\newblock Fast analytical methods for macroscopic electrostatic models in
  biomolecular simulations.
\newblock {\em SIAM Rev. 53\/} (2011), 683--720.

\bibitem{XCY:JCP:11}
{\sc Xu, Z., Cheng, X., and Yang, H.}
\newblock Treecode-based generalized {Born} method.
\newblock {\em J. Chem. Phys. 134\/} (2011), 064107.

\bibitem{XLX:SIAP:13}
{\sc Xu, Z., Liang, Y., and Xing, X.}
\newblock Mellin transform and image charge method for dielectric sphere in an
  electrolyte.
\newblock {\em SIAM J. Appl. Math. 73\/} (2013), 1396--1415.

\bibitem{Ying:SCM:12}
{\sc Ying, L.}
\newblock A pedestrian introduction to fast multipole methods.
\newblock {\em Sci. China Math. 55\/} (2012), 1043--1051.

\bibitem{YBZ:JCP:04}
{\sc Ying, L., Biros, G., and Zorin, D.}
\newblock A kernel-independent adaptive fast multipole algorithm in two and
  three dimensions.
\newblock {\em J. Comput. Phys. 196\/} (2004), 591--626.
\end{thebibliography}
\end{document}